\documentclass[
reprint,
superscriptaddress,
 showpacs,preprintnumbers,
 bibnotes,
 amsmath,amssymb,
 aps,
 prb,
]{revtex4-1}

\usepackage{graphicx}
\usepackage{dcolumn}
\usepackage{bm}
\usepackage{epstopdf}

\usepackage[utf8]{inputenc}
\usepackage{color}
\usepackage{float}
\usepackage{amsfonts}

\begin{document}


\title{All-electric single electron spin-to-charge conversion}
\author{J. Paw\l{}owski}
\email[]{jaroslaw.pawlowski@pwr.edu.pl}
\affiliation{
Department of Theoretical  Physics, Faculty of Fundamental Problems of Technology, Wroc\l{}aw University of Science and Technology, Wybrze\.{z}e Wyspia\'{n}skiego 27, 50-370 Wroc\l{}aw, Poland}
\author{G. Skowron}
\affiliation{
	Faculty of Physics and Applied Computer Science,
	AGH University of Science and Technology, Krak\'{o}w, Poland}
\author{M. G\'{o}rski}
\affiliation{
	Faculty of Physics and Applied Computer Science,
	AGH University of Science and Technology, Krak\'{o}w, Poland}
\author{S. Bednarek}
\affiliation{
	Faculty of Physics and Applied Computer Science,
	AGH University of Science and Technology, Krak\'{o}w, Poland}

\date{\today}

\begin{abstract}
We examine spin-dependent displacement of a single electron, resulting in separation and relocation of the electron wavefunction components, and thus charge parts, corresponding to opposite spins. This separation is induced by a pulse of an electric field which generates varying Rashba type spin-orbit coupling. This mechanism is next implemented in a nanodevice based on a gated quantum dot defined within a quantum nanowire. The electric field pulse is generated by ultrafast changes of voltages, of the order of several hundred mV, applied to nearby gates. The device is modeled realistically with appropriate material parameters and voltages applied to the gates, yielding an accurate confinement potential and Rashba coupling. At the end, we propose a spin-to-charge conversion device, which with an additional charge detector will allow for electron spin state measurement.
\end{abstract}


\maketitle


\section{Introduction}

There is currently great interest in control and manipulation of individual electrons in semiconductor nanostructures \cite{hanson}.
Such systems have a large variety of applications in fields such as modern fast electronics, spintronics\cite{zutic} and recently valleytronics\cite{valley},
which involves the so-called valley degree of freedom of an electron (present e.g., in hexagonal monolayers like graphene, bismuthene or MoS$_2$).
Also, various fundamental quantum-related phenomena 
can be studied using such systems\cite{splatanieeksperymenty1,splatanieeksperymenty2,winelandprzeglad}.
These studies may include topological effects \cite{hasan,nori}, recently introduced exotic quasi-particles \cite{cz1,cz2,cz3,cz4}, or implementation of quantum computation.

The electron qubit may be implemented in several ways\cite{owen}: as a charge qubit \cite{chq1,chq1a,chq2,chq2a,chq3,chtos}, a spin qubit\cite{loss, loss1} 
or encoded in electronic Schr\"odinger cat states\cite{koty1,koty2}. 
The solid-state qubit based on electron spin in electrostatic quantum dots is the easiest to implement and is one of a few promising candidates for quantum computing\cite{qrew}. 
The Rashba type spin-orbit coupling (RSOI) \cite{rashba, winkler}, which couples orbital and spin degrees of freedom, allows for efficient manipulation of the spin qubit \cite{nowack, np0, japarize, lopez, sherman, sherman1, ramsak, szaszko, pawlowski1, pawlowski2}. Moreover, there are several possibilities to obtain a scalable quantum computation architecture consisting of multiple electron spins. They involve capacitive\cite{8,coupled,cp3,5}, exchange\cite{cp5,cp4,10,12} or hybrid\cite{cp1,cp2,6,7,9} coupling of such qubits into registers, opening a way towards universal two qubit operations. 

Aside from the ability to perform operations, we also need to be able to initialize and read out the qubit state after the operation has been done\cite{11}. While for the charge qubit readout we can use quantum point contacts (QPCs) \cite{19,hanson,qpc} as charge detectors\cite{cd1,cd2,17a}, the spin-qubit requires the prior spin-to-charge conversion step\cite{21,edel}, typically employing the Pauli spin blockade\cite{nowack,np0,21,pauli} or spin-selective tunneling rates\cite{20}. Spin selection may additionally exploit metastable excited charge states\cite{21}, or the inverse Edelstein effect\cite{edel}. Likewise, the singlet-triplet qubit is read out by mapping these states onto different charge states\cite{17,18,6}.

The passive readout performed using spin-selective tunneling to nearby leads
is, unfortunately, relatively slow.
We propose a different approach, based on separation of a single electron wavefunction into two parts of definite spin, resulting effectively in spin-to-charge conversion. 
The process is done all-electrically via the RSOI, without the use of magnetic fields or optical transitions\cite{13,6}. 
A similar idea was presented by J. W\"{a}tzel \textit{et al.} in [\onlinecite{15}], but with very strong asymmetric pulses of an optically generated electric field, which in our proposal is created electrically by local gates. Moreover, in [\onlinecite{15}] momentum is generated using photons with a nonzero electric field component along the nanowire, while in our case the electric field is perpendicular to the wire, thus the problem of the electron tunneling outside the dot (along the wire) disappears.

As a result, our presented conversion scheme leads to ultrafast measurement\cite{nielsen} of the single electron spin. This implementation conforms to a new sub-discipline, the spin-orbitronics, where spin generation, manipulation and detection are performed solely by electrical means through the RSOI\cite{nitta}.

\section{Device model and calculation method}

\begin{figure}[h]
\includegraphics[width=8.0cm]{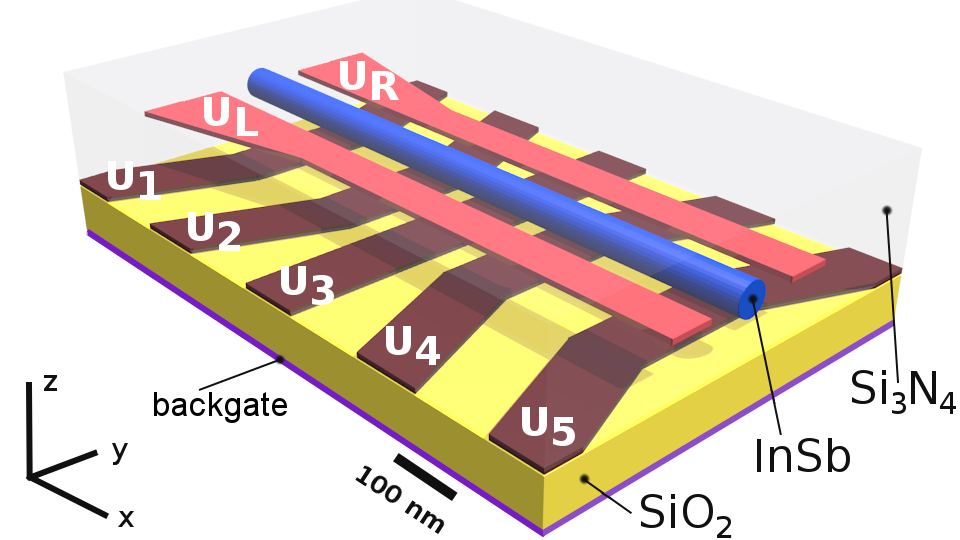}
\caption{\label{fig:1} The modeled nanodevice consists of an InSb nanowire and a layout of nearby gates. The bottom gates are used to create a confinement potential in the $x$-direction and, in the second stage of conversion, to generate a potential barrier separating previously split opposite spin densities. The lateral gates serve to generate a spin-orbit pulse within the wire, splitting spin densities spatially.}
\end{figure}

The modeled nanodevice is made of a catalytically grown InSb quantum wire of a typical diameter of $50\,\mathrm{nm}$ placed on a system of 5 bottom gates $U_{1...5}$, as shown in Fig.~\ref{fig:1}. At the sides of the wire two additional gates are placed, the left one $U_L$ and the right one $U_R$. The bottom and lateral gates are spatially separated from the wire by an insulating layer of Si$_3$N$_4$, thus minimizing the leakage current from the wire\cite{phd} (see appendix for additional information about the materials). The bottom gates are also separated from a strongly doped silicon substrate by an 80~nm layer of SiO$_2$. To the substrate, also serving as a backgate, we apply the reference voltage $V_0=0$. 

\begin{figure}[t]
	\includegraphics[width=2.8cm]{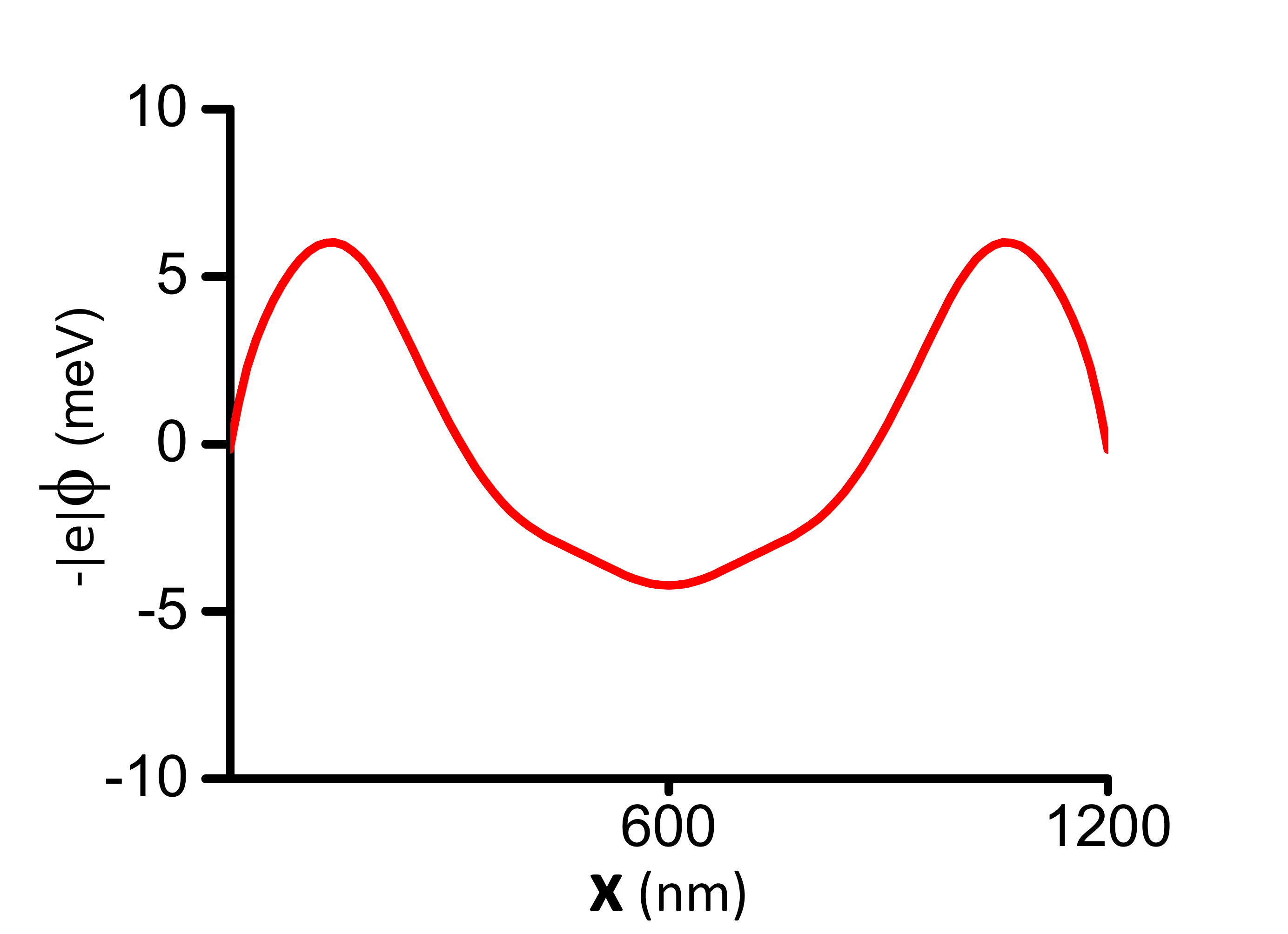}
	\includegraphics[width=2.8cm]{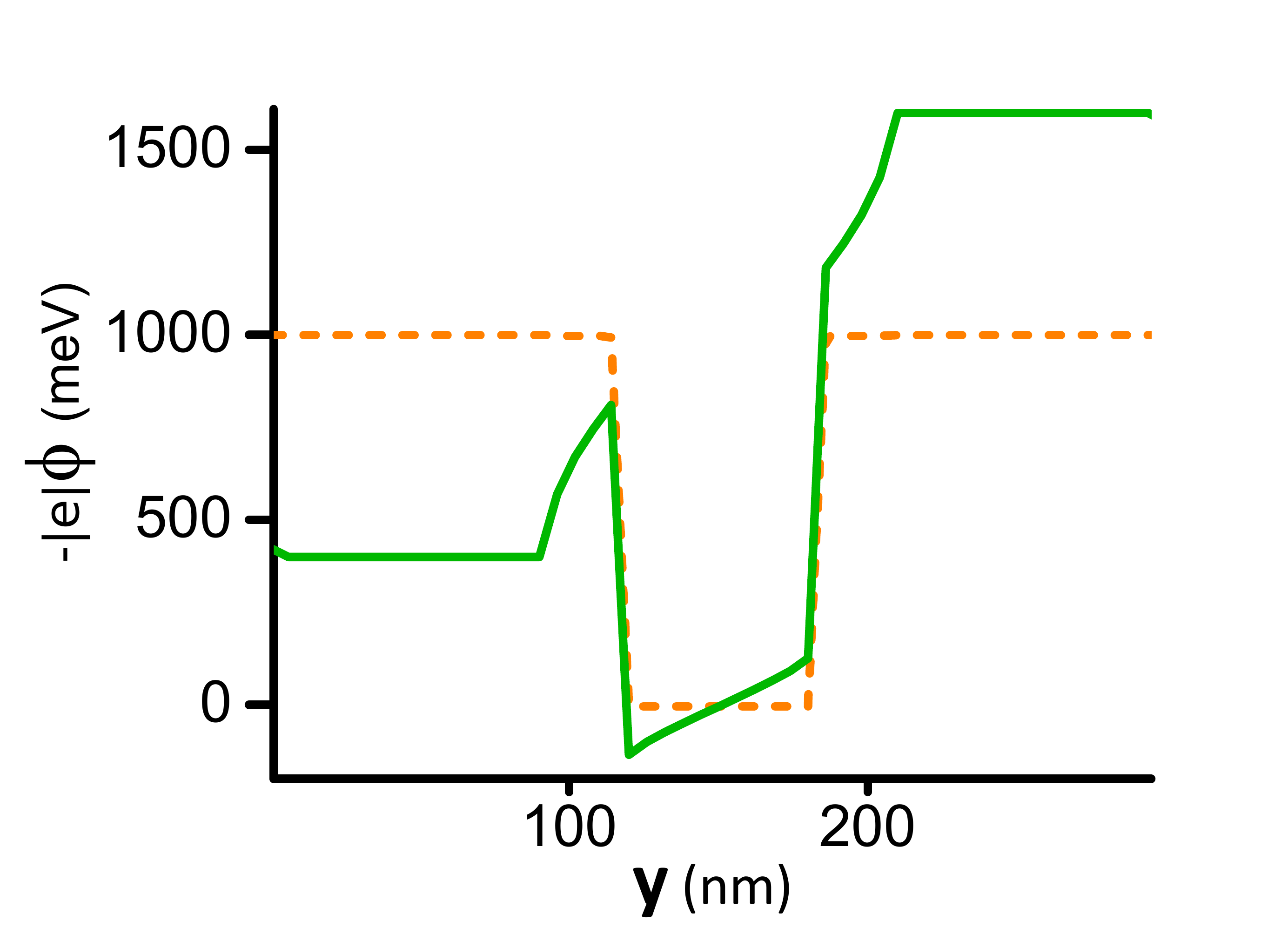}
	\includegraphics[width=2.8cm]{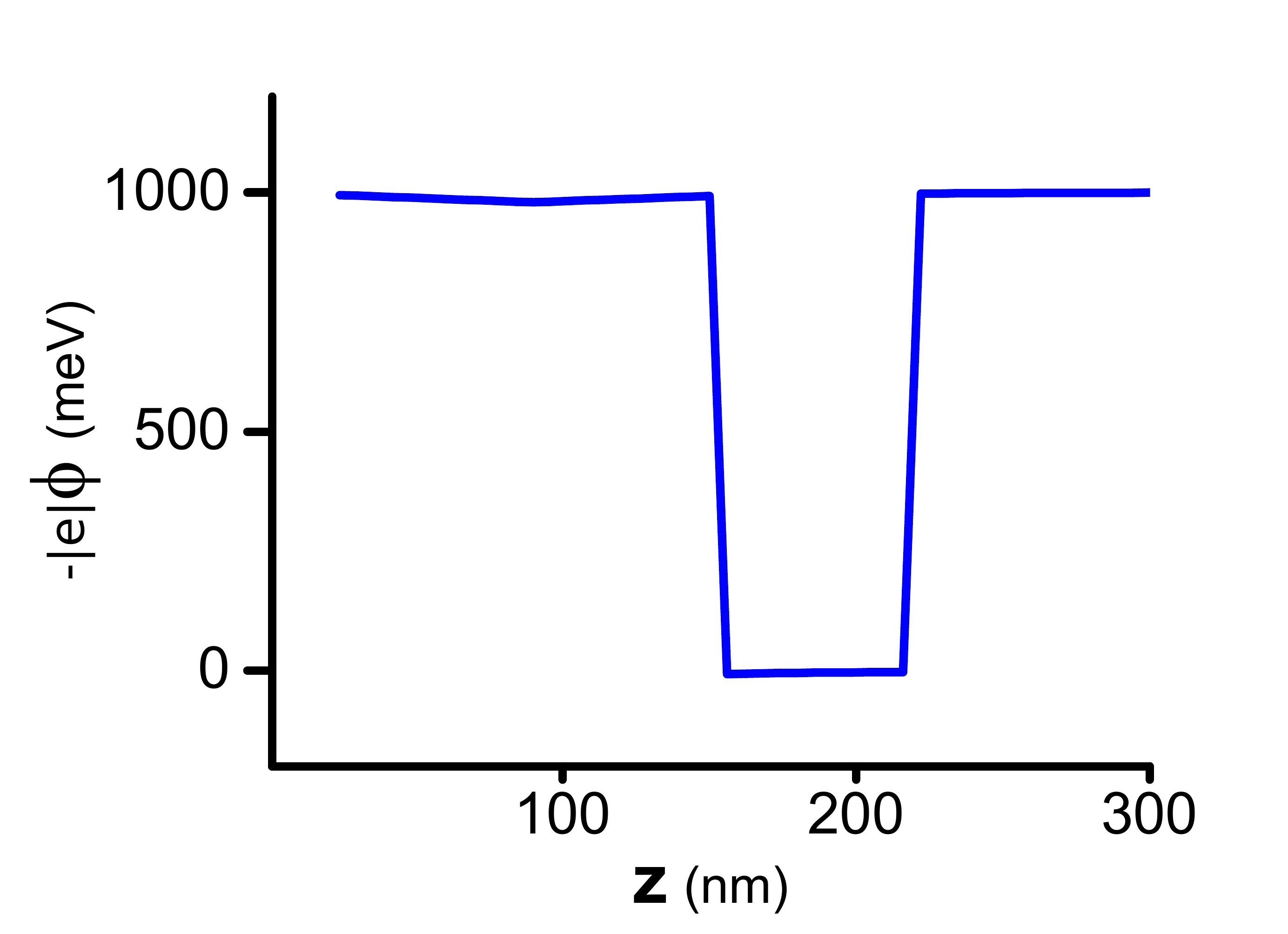}
	\caption{\label{fig:2}The confinement potential landscapes in the center of the wire during the first stage of conversion along the all three directions: (left) along the wire axis $x$, (middle) perpendicular to the wire along the $y$ direction connecting the lateral gates, and (right) along the $z$-direction, perpendicular to the substrate.} 
\end{figure}

The five bottom finger-like gates generate a confinement potential in the wire effectively forming a quantum dot which traps a single electron. The shape of this potential is shown in Fig.~\ref{fig:2}(left). Gate $U_3$, in the second stage of the device operation, is used to generate a barrier in the center of the wire, spatially separating and stabilizing the electron spin densities corresponding to opposite spin orientations. To gates $U_L$ and $U_R$ we apply a voltage pulse, generating a lateral electric field, visible (when the field is maximum) in Fig.~\ref{fig:2}(middle).

The time-dependent Hamiltonian of a single electron inside the wire nanostructure, aside from its kinetic term, contains the quantum dot potential $\phi(\mathbf{r},t)$ controlled by voltages applied to the bottom gates:
\begin{equation}
H(\mathbf{r},t)=\left(-\frac{\hbar^2}{2m}\nabla^2 - |e|\phi(\mathbf{r},t)\right)\mathbf{1}_2+H_R(\mathbf{r},t),
\label{ham}
\end{equation}
with the InSb band mass $m=0.014\,m_e$. Additionally the key element for the conversion method is the presence of the RSOI, which is manipulated by the perpendicular electric field $E_y$ created using the lateral gates.

The general RSOI Hamiltonian accounting for an inhomogenous electric field $\mathbf{E}$ is given by 
$H_R(\mathbf{r},t)=\frac{\gamma_\mathrm{3D}|e|}{\hbar}(\mathbf{E}(\mathbf{r},t)\times\mathbf{p})\cdot\boldsymbol\sigma$
with the space dependent electric field $\mathbf{E}(\mathbf{r},t)=-\boldsymbol\nabla \phi(\mathbf{r},t)$ (for InSb $\gamma_\mathrm{3D}=5.23$~nm$^2$ [\onlinecite{winkler}]), the momentum operator $\mathbf{p}=-i\hbar\boldsymbol\nabla$, $\boldsymbol\nabla \equiv \left[\partial _x,\partial _y,\partial _z\right]$, and the vector of Pauli matrices: $\boldsymbol\sigma \equiv \left[\sigma _x,\sigma _y,\sigma _z\right]$. 
There are two contributions to the electric field within the wire: first, generated by the lateral gates, and second a much smaller field that generates the confinement potential along the wire.
The key electronic behavior will be its spin-dependent motion along its only degree of freedom, that is, the $x$ axis. 
This clearly shows that the greatest contribution to the RSOI Hamiltonian is introduced by the term $-\frac{\gamma_{3D}|e|}{\hbar}E_yp_x\sigma_z$, coupling the spin $z$-component $\sigma_z$ with the electron momentum $p_x$ along the wire.
The asymmetry of the crystallographic structure inducing the Dresselhaus spin-orbit interaction vanishes if the nanowire is grown along the [111] crystallographic direction\cite{nodrr,winkler}.

To simulate the nanodevice operation we solve numerically the time-dependent Schr\"odinger equation with the potential $\phi(\mathbf{r},t)$ and the electric field $\mathbf{E}(\mathbf{r},t)$, both obtained from Poisson equation.
The solution of the time-dependent Schr\"odinger equation with the Hamiltonian (\ref{ham}) has a two-row spinor form  $\Psi(\mathbf{r},t)=\left(\psi_\uparrow(\mathbf{r},t),\psi_\downarrow(\mathbf{r},t)\right)^T$, with spin-up and spin-down wavefunction components.
The Poisson equation takes into account the time-dependent distribution of the electron density $|\Psi(\mathbf{r},t)|^2$ and the charge induced on material interfaces and the gates. It must thus be solved in a self-consistent manner along with Schrödinger's equation at every time step. Further details of the numerical method can be found in the appendix.

\section{Analytical considerations\\---single pulse}
In order to illustrate the basic concept of the proposed scheme and understand the influence of the time dependent RSOI on the ground state of the harmonic oscillator, we consider an effective 1D problem by freezing, in this paragraph, any motion in directions perpendicular to the wire: $y$, $z$. In the later part, where we discuss a particular proposal for realization of the nanodevice, we will return to full 3D calculations.

The one-dimensional form of Hamiltonian (\ref{ham}) for a single electron confined in a wire oriented along the $x$-axis takes the form
\begin{equation}\label{eq:ham1d}
h_{1D}(x)=(-\frac{\hbar^2}{2m}\partial_x^2 + u(x))1_2 + h_R,
\end{equation}
with the confinement potential $u(x)$ along the wire and the RSOI $h_R=-\frac{\gamma_{3D}|e|}{\hbar}E_y p_x\sigma_z$ induced by a perpendicular to the wire electric field $E_y$. 

If the confinement potential is parabolic $u(x)=m\omega_\mathrm{osc}^2x^2/2$ the electron set in motion oscillates (in a coherent state) with respect to the potential minimum with a frequency independent of the oscillation amplitude, thus behaves like a classical particle. By employing the Ehrenfest theorem we can describe the expectation value of position with classical equations of motion. We use $U_L$ and $U_R$ to generate the lateral time-varying electric field $E_y=E_y^0(1-\cos(\omega{}t))$ of angular frequency $\omega$ different from the harmonic potential eigenfrequency $\omega_\mathrm{osc}$. We rewrite the Hamiltonian (\ref{eq:ham1d}) into its classical form
\begin{equation}\label{anal}
 \mathcal{H}=\frac{p^2}{2m}+\frac{m}{2}\omega_\mathrm{osc}^2x^2-\varsigma F(1-\cos(\omega{}t))p,
\end{equation}
for $F=\frac{\gamma_{3D}|e|}{\hbar}E_y^0$ with $\varsigma=1$ for the upper spinor component $\psi_\uparrow$, and $\varsigma=-1$ for the lower component $\psi_\downarrow$. From the Hamilton equations we obtain an equation of motion for the expectation value of position:
\begin{equation}\label{newton}
\ddot{x}+\omega_\mathrm{osc}^2x=-\varsigma F\omega\sin(\omega{}t). 
\end{equation} 
This is the equation of the driven harmonic oscillator. For initial conditions $x(0)=0$ and $\dot{x}(0)=0$ we obtain the solution for spin-up component ($\varsigma=1$):
\begin{equation}\label{sol}
x_\uparrow(t)=\frac{F\omega}{\omega_\mathrm{osc}^2-\omega^2}\left(\frac{\omega}{\omega_\mathrm{osc}}\sin(\omega_\mathrm{osc}t)-\sin(\omega{}t)\right).
\end{equation} 
\begin{figure}[b]
	\includegraphics[width=7.9cm]{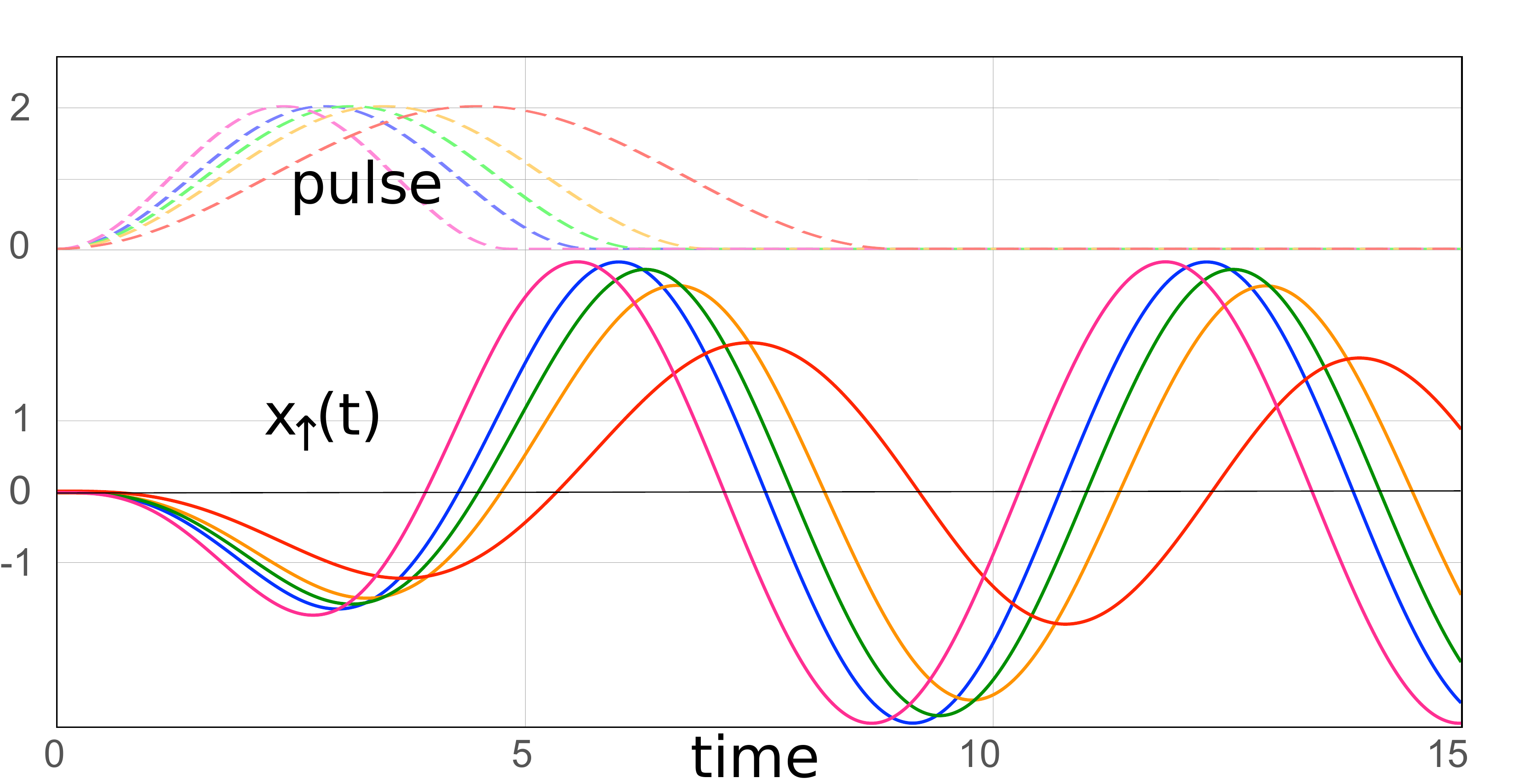}
	\caption{\label{fig:3} The expectation value of position of the spin-up component $x_\uparrow(t)$ (solid lines) for a single driving pulse $F(1-\cos(\omega{}t))$ of different durations $T=2\pi/\omega$ (dashed lines). We set $F=1$ and $\omega=\{0.7,0.9,1.0,1.1,1.3\}$ for red, orange, green, blue and magenta curves respectively.} 
\end{figure}

To obtain the solution for a driving frequency equal to the eigenfrequency of the harmonic oscillator we take the limit $\omega\to\omega_\mathrm{osc}$, and get $x_\uparrow(t)=\frac{Ft}{2}\cos(\omega{}t)$. In the case of resonance, the amplitude of oscillations grows the fastest.
Now, let us note that since driving depends on the sign of $\varsigma$, the electron with spin-up will oscillate in the opposite direction to the electron with spin-down: $x_\uparrow(t)=-x_\downarrow(t)$. If the electron spin is not parallel to the $z$-axis, both spinor components $\psi_\uparrow$ and $\psi_\downarrow$ will move in opposite directions and oscillate in antiphase with growing amplitudes. Such spin-dependent oscillations induced by RSOI were used in [\onlinecite{ininjp}] for spatial separation of spin components in a planar heterostructure. 

Due to the fact that the nanodevice proposed in this paper consists of a nanowire surrounded with an insulator, we can apply stronger electric fields and separate the electron spins using a \textit{single pulse} of voltages. This lifts the requirement for resonant value of $T$. The only condition is an appropriately high pulse amplitude $F$ to facilitate spin separation. Fig.~\ref{fig:3} shows the solution of (\ref{newton}) for a single pulse $F(1-\cos(\omega t))$ lasting for $T=2\pi/\omega$. We observe sufficient spin separation for a very wide range of pulse durations, which translates into high immunity against non-optimal selection of $T$.

\section{Simulation results and discussion}
Let us now return to full 3D calculations.
\begin{figure}[b]
	\vspace{3mm}
	\includegraphics[width=8.1cm]{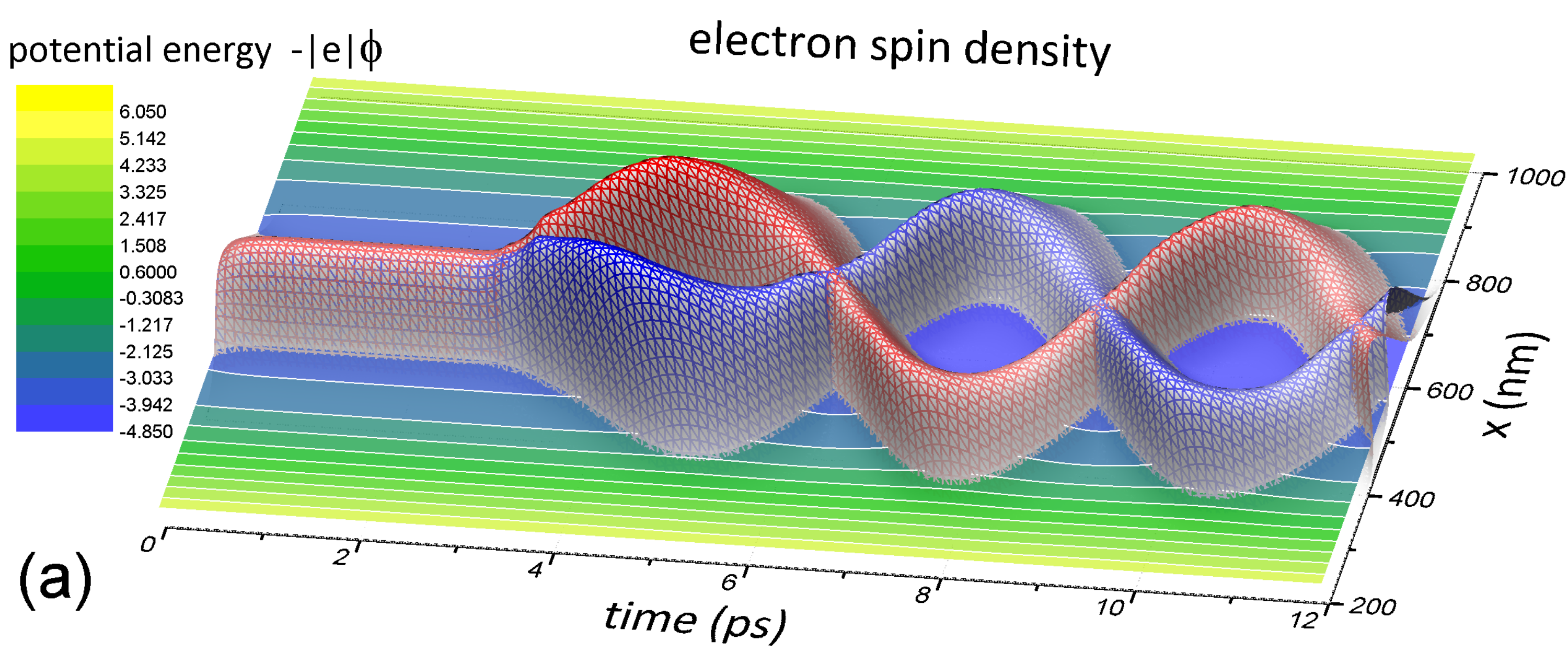}
	\includegraphics[width=3.7cm]{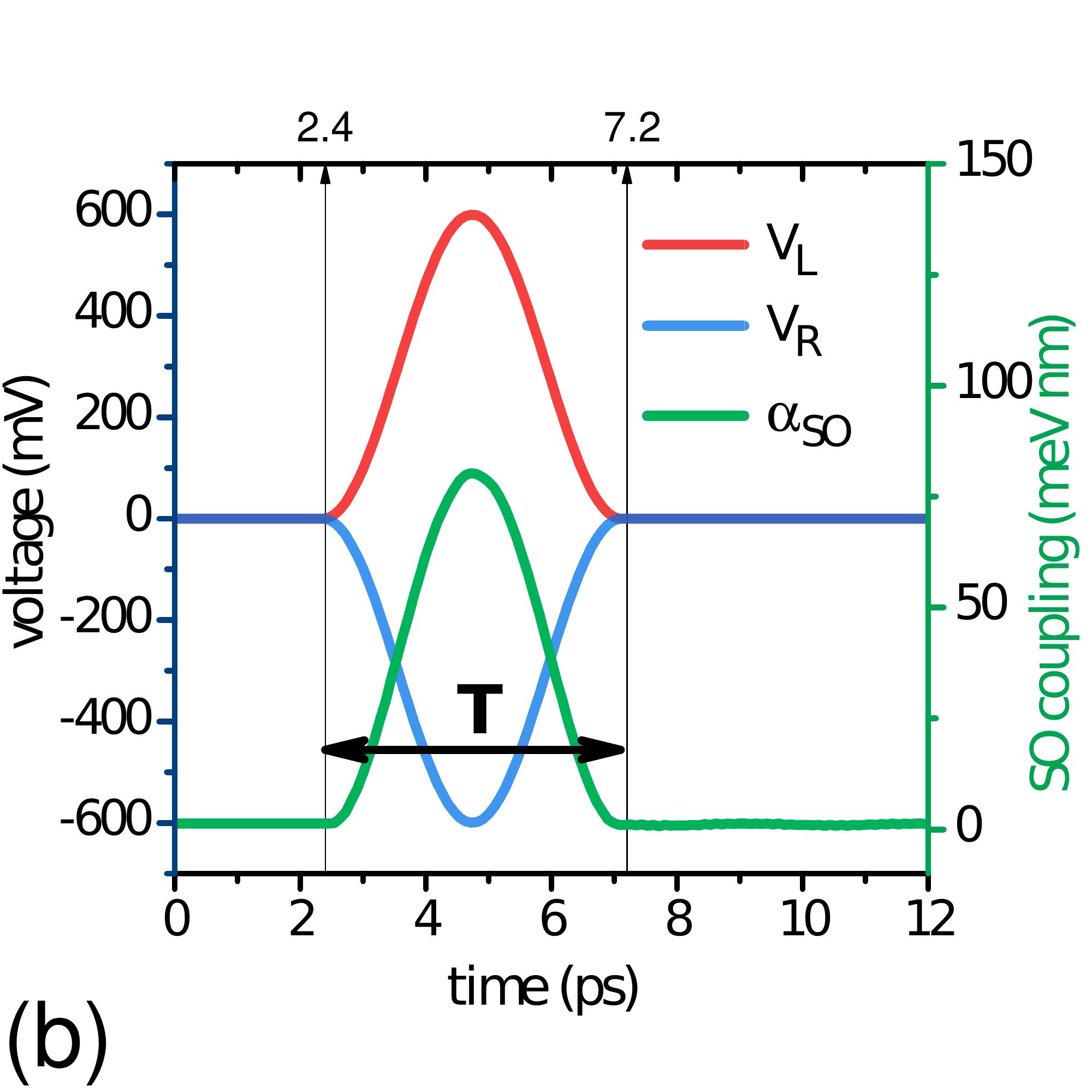}
	\includegraphics[width=3.7cm]{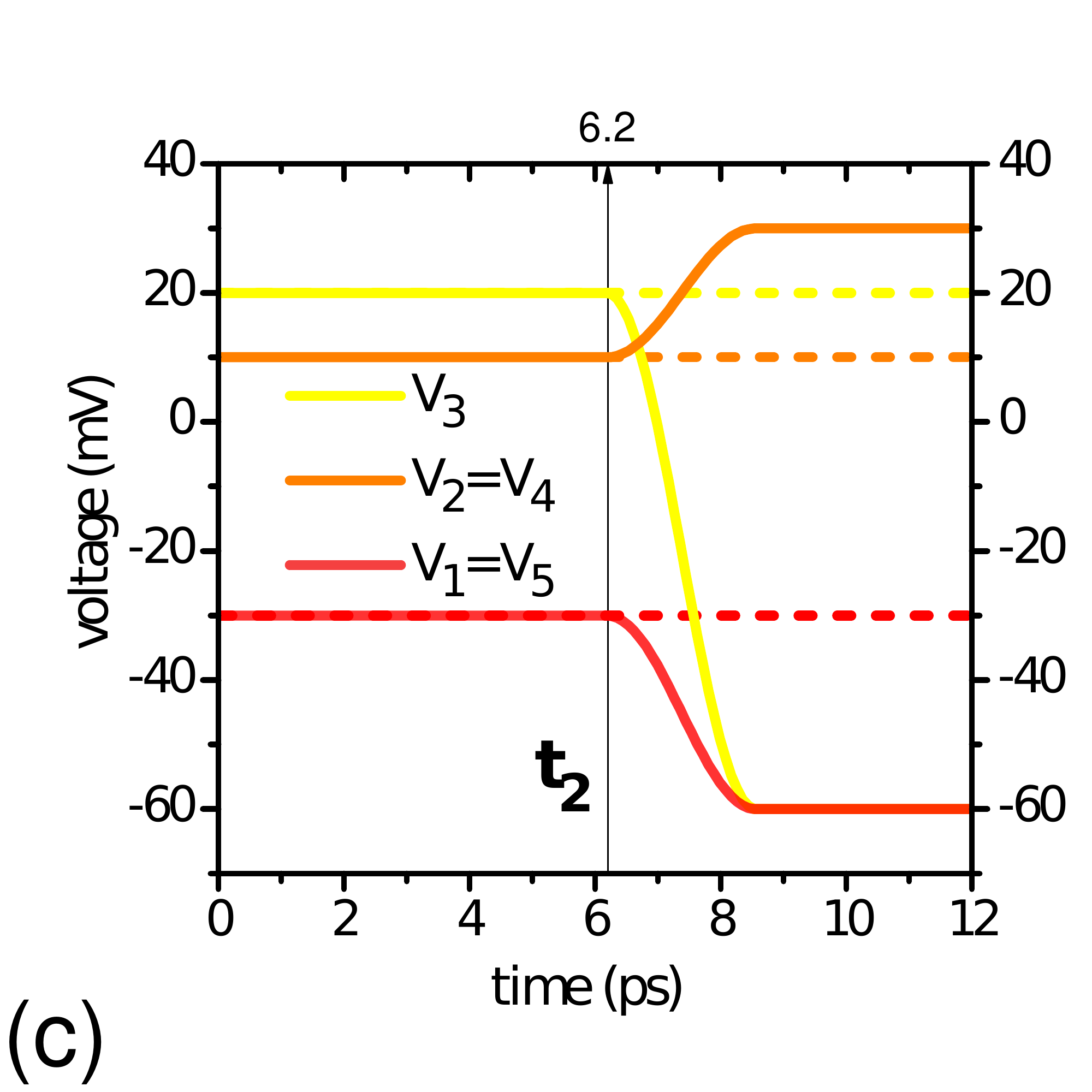}\vspace{2mm}
	\includegraphics[width=8.1cm]{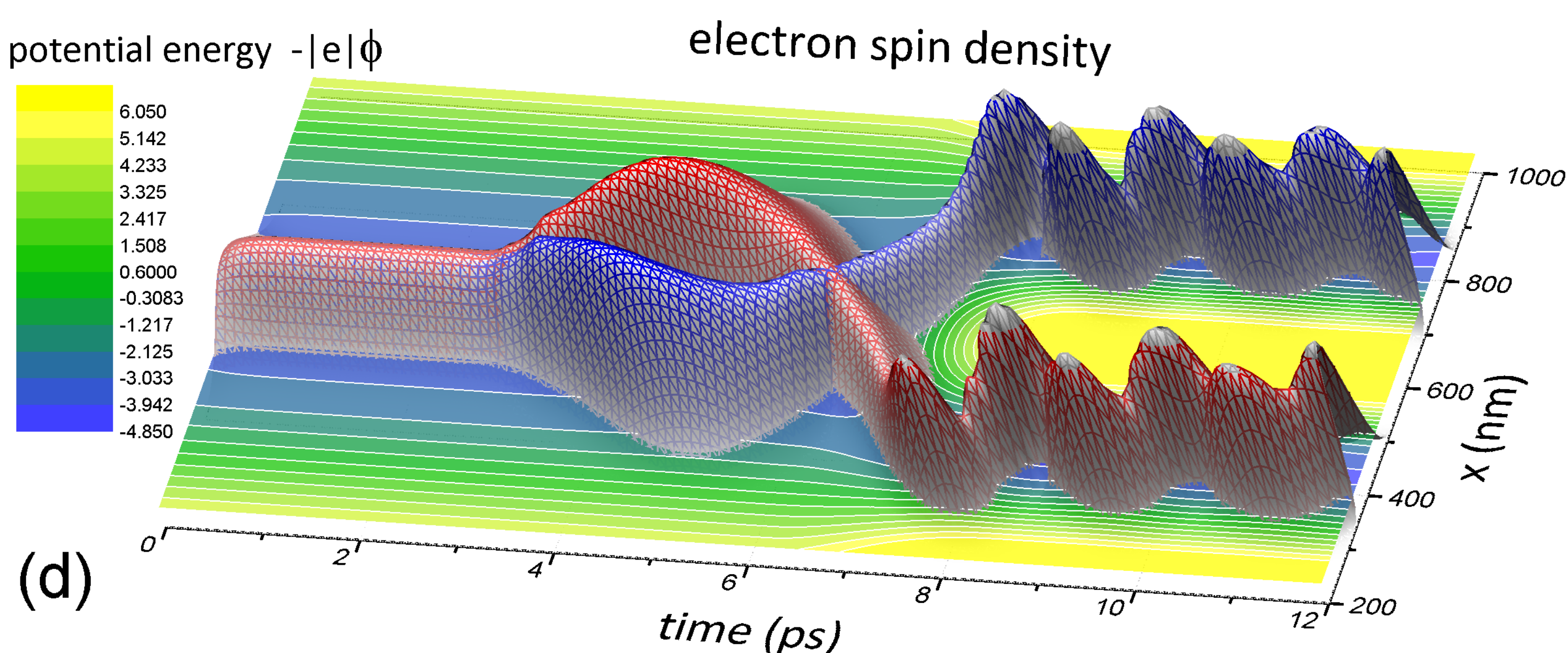}
	\caption{\label{fig:4} (a) The Rashba coupling pulse generates oscillations of the spin densities for spin-up (blue convex density) and spin-down (red convex density). They are in antiphase to each other. (b) Pulses of voltages on lateral gates induce an RSOI pulse. (c) At the second stage of conversion we set up a barrier in the middle of the wire to separate and stabilize spatially both spin densities. (d) Simulations with an additional barrier rise shown. Changes to the confinement potential are shown as a color-map beneath the densities.}
\end{figure}
In Fig.~\ref{fig:4} we presented a time evolution of the spin density with spin initially aligned along the $y$ direction. At $t_1=2.4$~ps we turn on a \textit{cosine-like} pulse of voltages $V_{0L(R)}(1-\cos(\omega t))$ applied to lateral gates $U_L$ i $U_R$ of amplitudes $V_{0L}=600$~mV, $V_{0R}=-600$~mV and duration $T=4.5$~ps. The time courses of $V_{L,R}$ are shown in Fig.~\ref{fig:4}(b). The electric field induced by the voltage pulse generates a pulse of RSOI of the same duration within the wire, marked as the green curve. An increase and subsequent decrease of the RSOI coupling causes spatial separation of both spinor components. This way, spin densities corresponding to opposite spins: spin-up $\rho_\uparrow$ (blue in Fig.~\ref{fig:4}(a)) and spin-down $\rho_\downarrow$ (red) start oscillating in antiphase. The amplitude of these oscillations (up to certain extent) is proportional to the amplitude of the RSOI pulse. The spin density for the upper spinor component, namely spin-up is calculated as: $\rho_\uparrow(x,t)=\int dydz |\psi_\uparrow(\mathbf{r},t)|^2$ and similarly $\rho_\downarrow(x,t)$ for $|\psi_\downarrow(\mathbf{r},t)|^2$.

After the spatial separation of the electron densities of opposite spins, we raise a barrier in the middle of the wire to separate both wavepacket parts permanently and to stabilize them in the left and right halves of the wire. We take an assumption that the separating barrier is not set up instantaneously and its rise rate equals half of the pulse duration, namely $T/2$. To do it, we lower $V_3$ voltage at $t_2=6.2$~ps and tune the remaining voltages applied to bottom gates. As a result, we get a potential barrier in the center of the wire along the $x$-direction. The barrier is shown in yellow for the second part of the simulation in Fig.~\ref{fig:4}(d). Voltages applied to bottom gates $V_{1\dots5}$ initially equal $-30,10,20,10,-30$~mV. After we set up the barrier they change to $-60,30,-60,30,-60$~mV, as marked in Fig.~\ref{fig:4}(c). If the barrier is set up at the right moment, the spin densities cease to oscillate and eventually stabilize inside the left and right parts of the wire, as shown in Fig.~\ref{fig:4}(d).

The initial spin in simulations shown in Fig.~\ref{fig:4} is set in parallel to the $y$-axis, thence it is an equally weighted linear combination of spin-up and spin-down:
$\frac{1}{\sqrt{2}}\left(|\uparrow\rangle + i|\downarrow\rangle\right)$.
That is why the conversion yields equally distributed charge inside both (left and right) parts of the wire $Q_L=Q_R=0.5$. The charge in both sides is obtained by integrating the total electron density in either left or right half of the wire. For the left: $Q_L(t)=\int_0^{l/2}d^3r\,\Psi^\dag\Psi=\int_0^{l/2} d^3r\, (|\psi_\uparrow(\mathbf{r},t)|^2+|\psi_\downarrow(\mathbf{r},t)|^2)$, where $l=1\;\mu$m is the wire length (note the integral limits). We proceed in a similar way to get the total charge on the right side: $Q_R(t)=\int_{l/2}^l d^3r\,\Psi^\dag\Psi$.

\begin{figure}[t]
	\includegraphics[width=7cm]{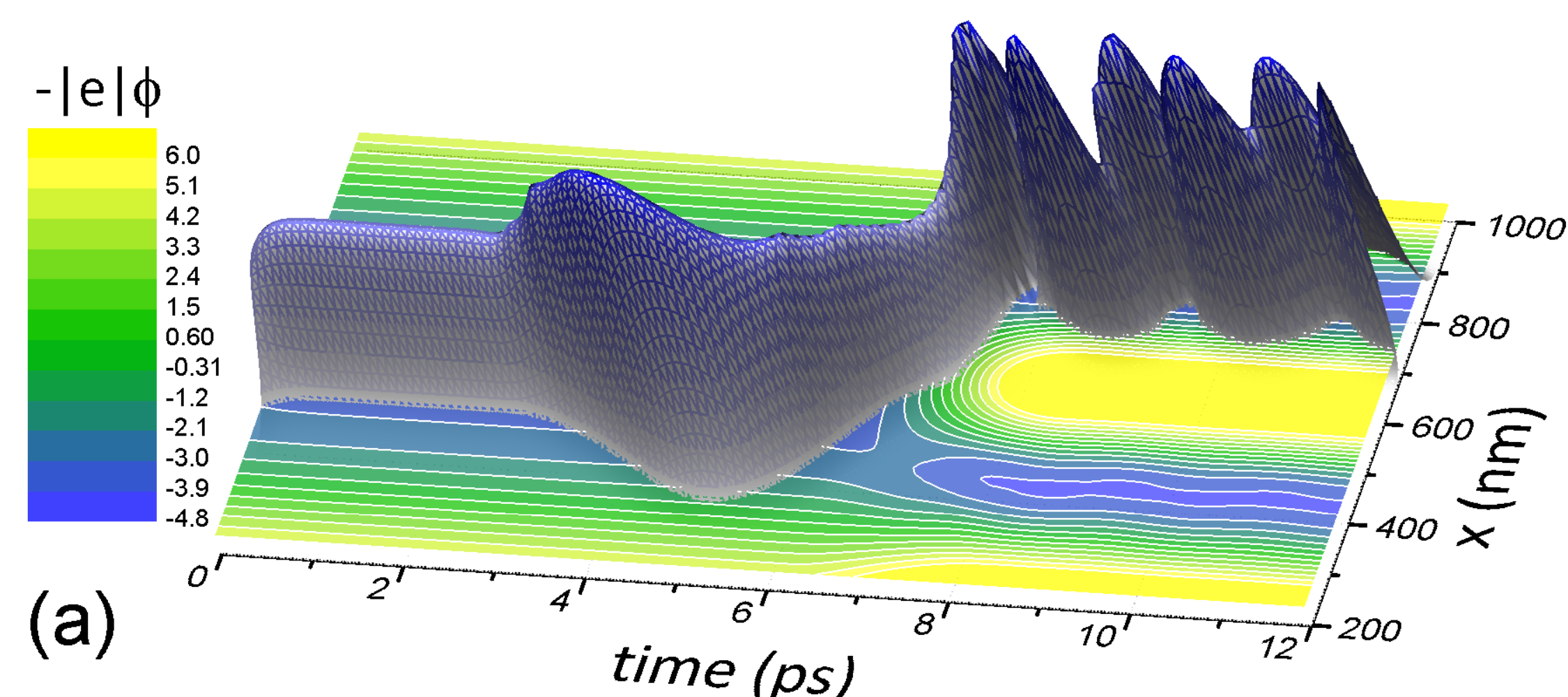}
	\includegraphics[width=7cm]{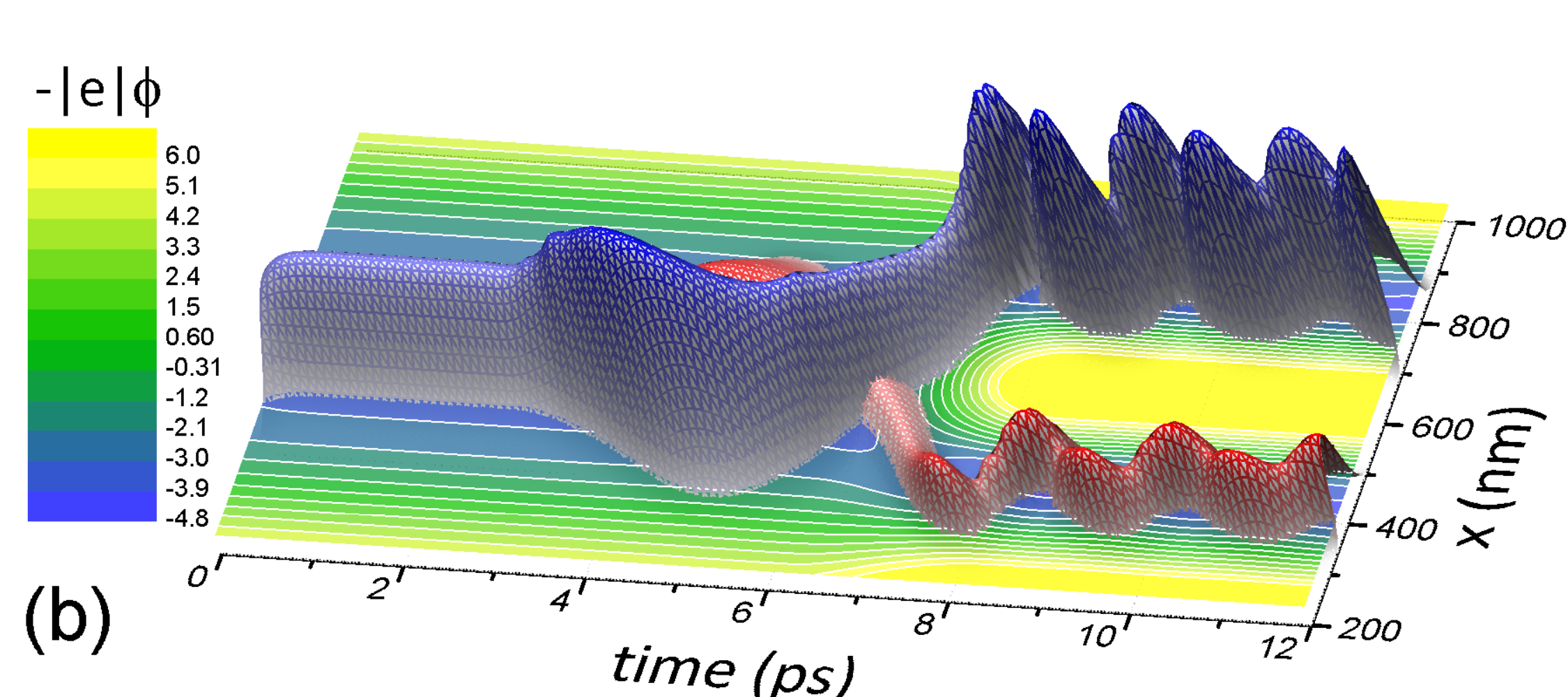}
	\includegraphics[width=7cm]{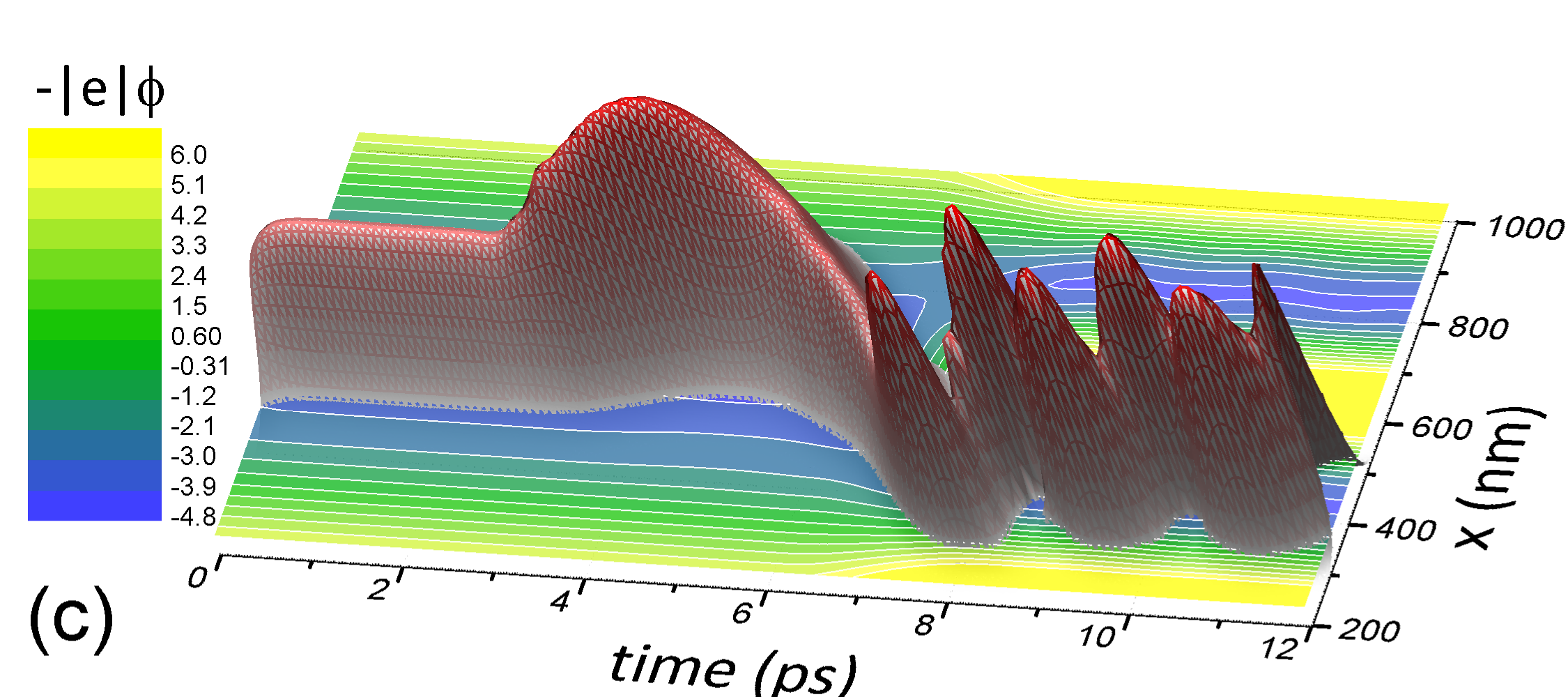}
	\caption{\label{fig:5} Conversion for various initial spin setups: (a) spin-up orientation, (b) orientation between spin-up and spin-down, (c) and spin-down. After conversion the sides contain charge $Q_R$ and $Q_L =1-Q_R$ proportional to the initial spin contributions.}
\end{figure}

In case of a non-equally weighted linear combination of spin-up and spin-down, the same proceeding would lead us to a different final charge distribution. In Fig.~\ref{fig:5}(a) we see a result of conversion performed for spin initially oriented along the $z$-axis (namely spin-up), and in Fig.~\ref{fig:5}(c) we see the same for spin-down. In the first case, at the end, we get $Q_R\simeq1$, and in the latter $Q_R\simeq0$. For an intermediate situation with spin tilted away from the spin-up orientation by $\vartheta=\pi/4$ we get a non-equal charge distribution, with a greater amount of charge on the right side of the wire (Fig.~\ref{fig:5}(b)).

\begin{figure}[t]
	\includegraphics[width=8.7cm]{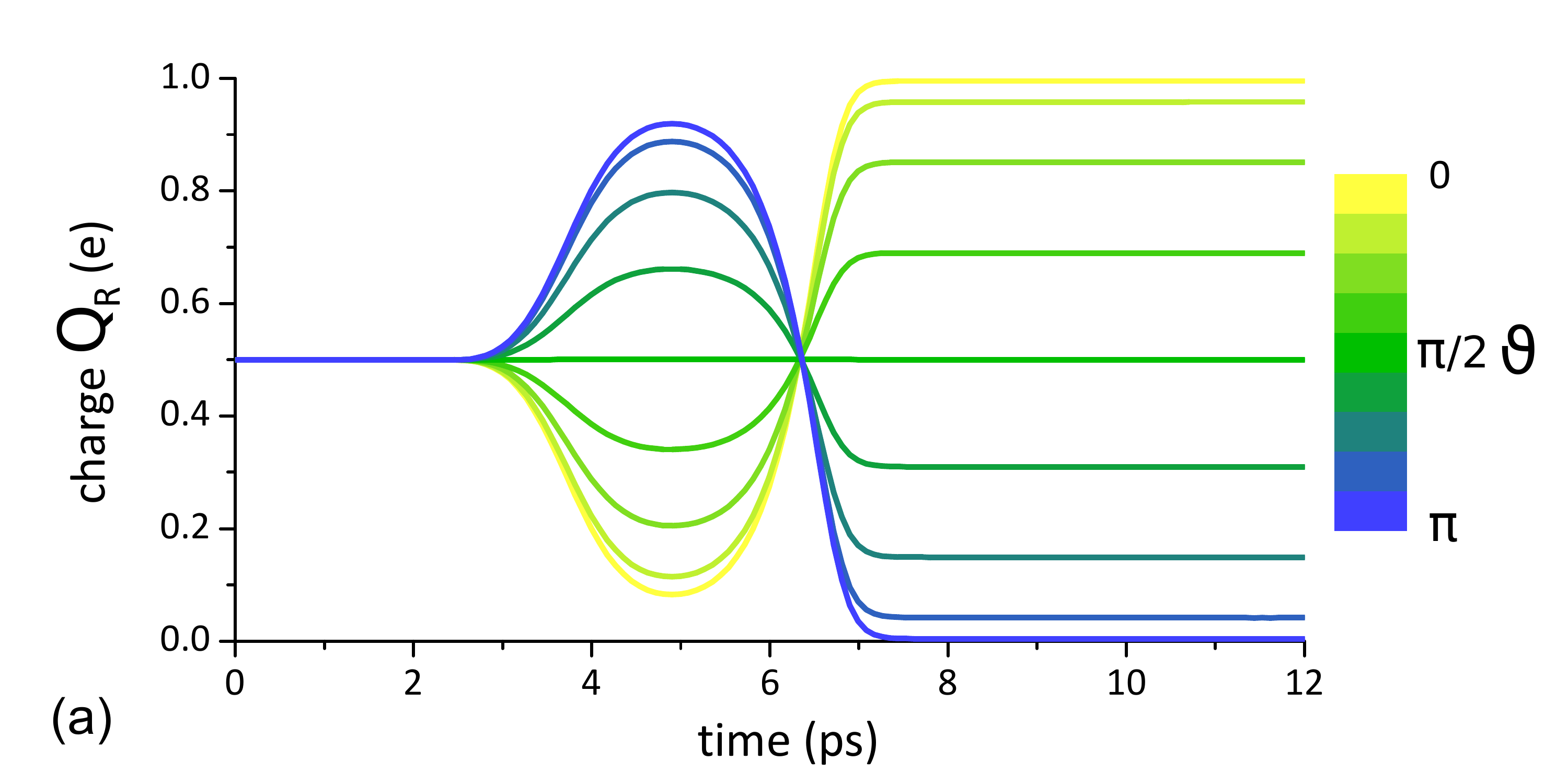}
	\includegraphics[width=4.7cm]{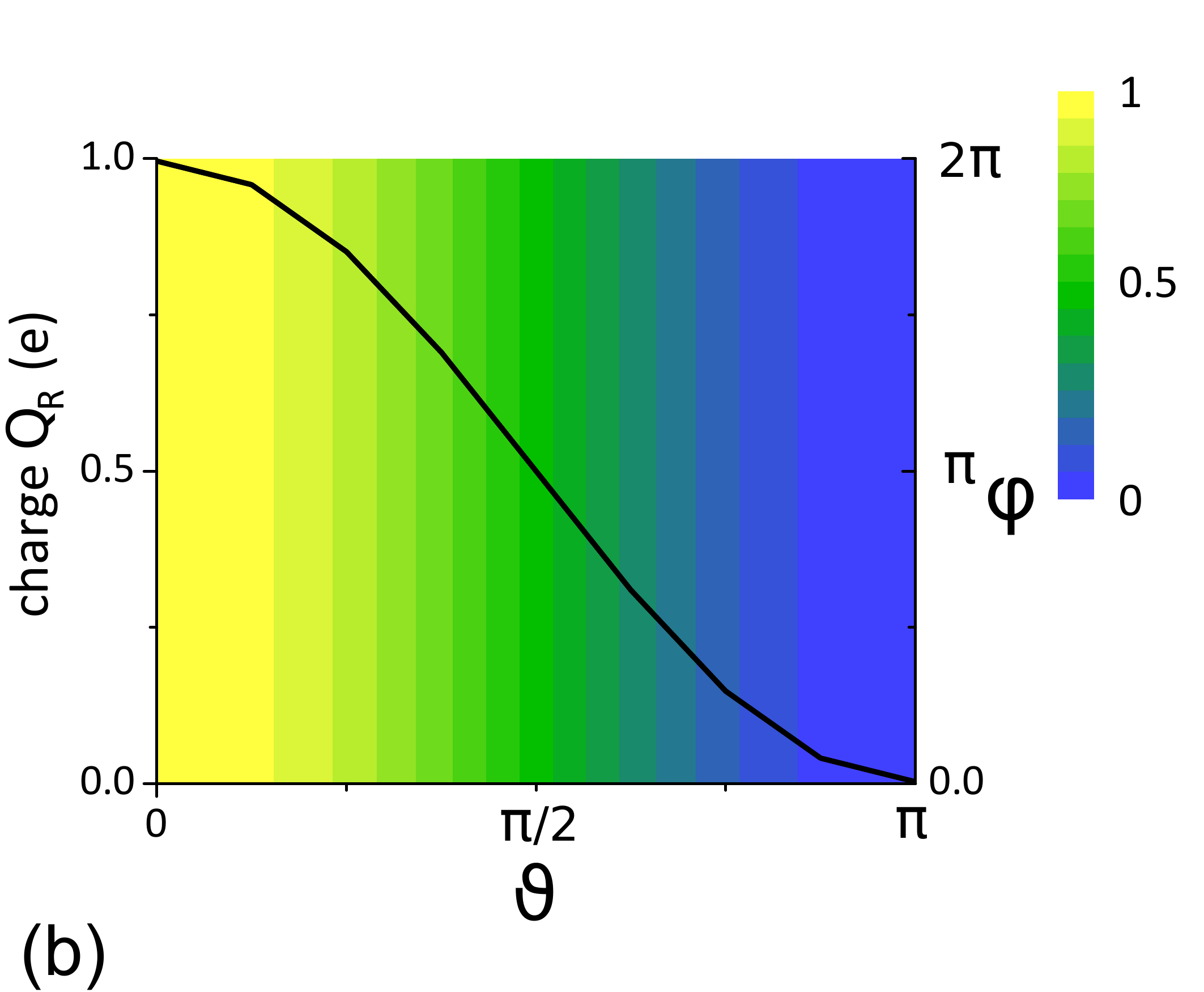}
	\includegraphics[width=3.8cm]{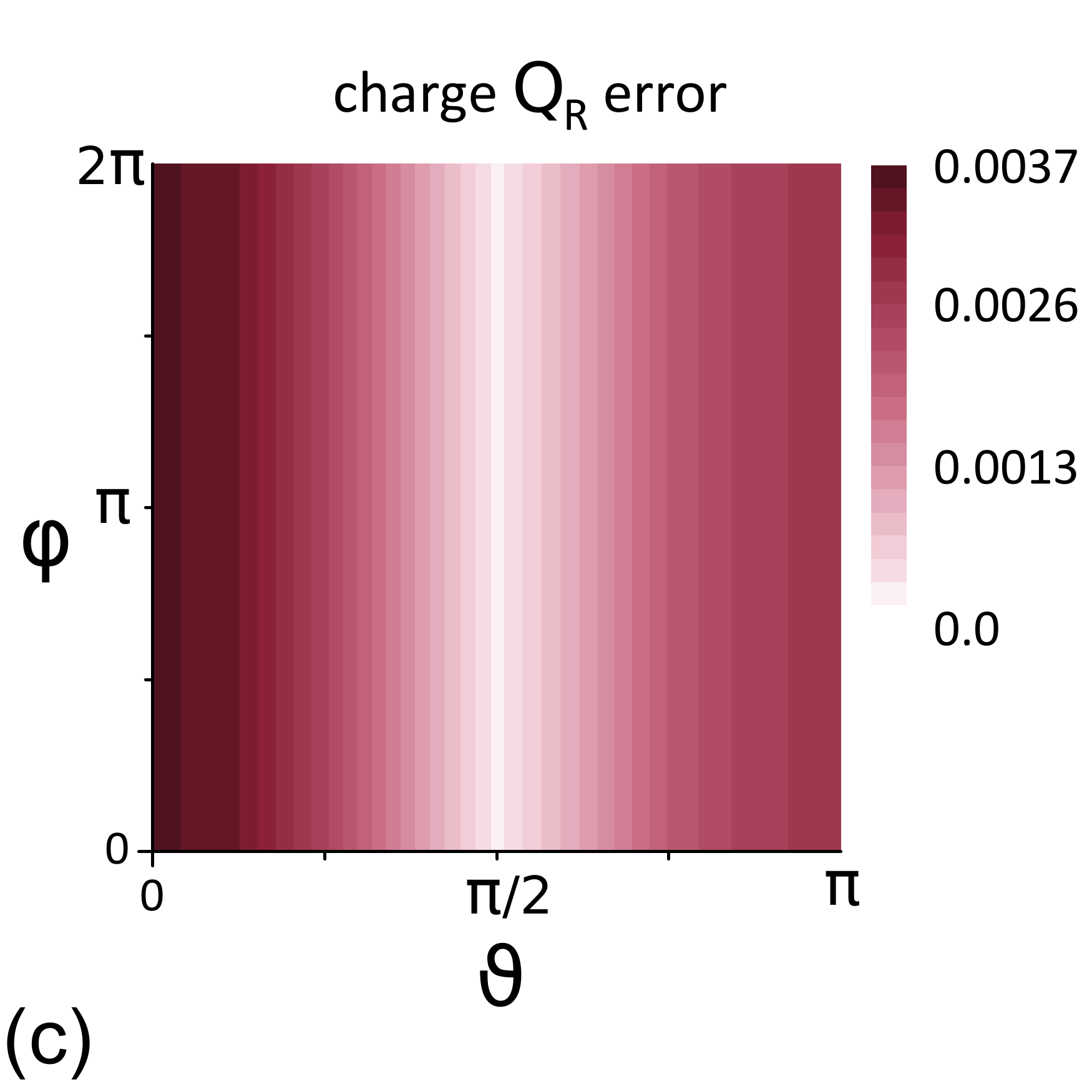}
	\caption{\label{fig:6} Spin-to-charge conversion: (a) time course of the charge amount in the right half of the wire $Q_R(t)$ depending on the initial spin orientation. The spin is parametrised by angles $(\vartheta,\varphi)$ on the Bloch sphere. (b) Final charge in the right half depends on the spin $z$-projection, thus on the angle $\vartheta$, yet is independent of $\varphi$. (c) Error of conversion depending on the initial spin orientation.}
\end{figure}

Let us look at Fig.~\ref{fig:6}(a), at the course of $Q_R(t)$ during time evolution for various initial spin orientations, i.e. configurations of the spin part of spinor $\Psi$: $\left(\cos(\frac{\vartheta}{2}),e^{i\varphi}\sin(\frac{\vartheta}{2})\right)^T$, parametrized by $\vartheta \in [0,\pi]$ and $\varphi \in [0,2\pi)$ on the Bloch sphere.
We observe a gradual decrease of the final charge $Q_R$ in the right side as the azimuthal angle $\vartheta$ increases.
In general, the method quite consistently yields amplitudes of finding the electron inside the right and left halves of the wire close to $\cos(\frac{\vartheta}{2})$ and $e^{i\varphi}\sin(\frac{\vartheta}{2})$ respectively.
Therefore we obtain the probability, and effectively charge depicted by color map in Fig.~\ref{fig:6}(b), nearly $\cos^2(\frac{\vartheta}{2})$ for $R$ side, 
and similary $\sin^2(\frac{\vartheta}{2})$ for $L$ side (not presented).
As expected, charge $Q_R$ and $Q_L$ do not depend on the polar angle $\varphi$.

Now one can perform a measurement of charge trapped in the left (right) half of the device using a nearby electric field sensor such as a QPC\cite{20,qpc,qpc1,qpc2}.

\subsection*{Parameters tuning and conversion fidelity}
The conversion method introduces a small error, which we estimate for the presented structure. In a physical nanodevice the confinement potential deviates from parabolicity. This affects spin densities and they deviate from Gaussians (typical for coherent states) and now possess small tails (which leaves some density on the other side of the wire). Also, nonparabolicity affect the amplitude of their oscillations. This amplitude should be large enough to facilitate full spatial separation after setting up a potential barrier in the middle. Fortunately, the amplitude can be almost arbitrarily increased by applying higher voltage pulses to $V_{0L}$ and $V_{0R}$ Similarly, as in the analytic solution (\ref{sol}), increasing the pulse strength $F$ proportionally increases the displacement of spin densities. 
Dynamic parameters, namely the pulse duration $T$, and $t_2$---the moment the separating barrier starts to rise, should also be properly tuned. $T$ should assume a value giving the highest packet’s displacement amplitude (for given $V_{0L,R}$) and $t_2$ a value so that to separate spin densities when they are the furthest away from each other during separation in the second stage of conversion.

By comparing resulting final charge $Q_R$ in the right half (the probability of finding the electron in the right half) with the (ideal) probability of finding the electron with spin-up in the initial state $\cos^2(\frac{\vartheta}{2})$, we obtain the error of the conversion method:
\begin{equation}\label{eq:ham1d}
\mathrm{error}(\vartheta,\varphi)=\cos^2(\vartheta/2)-Q_R(\mathrm{t_{end}}).
\end{equation}
In Fig.~\ref{fig:6}(c) we see that for optimally chosen parameters in the presented simulation i.e., $T=4.5$~ps and $t_2=6.2$~ps we get a conversion error below $0.4$\% ($0.3$\%), giving the greatest values for spins set in parallel (antiparallel) to the $z$-axis. Just like the charge $Q_R$, the error turns out to be also independent of the angle $\varphi$.

The optimal duration $T$ of the pulse depends on the shape of the driving pulse and has a value between $T_\mathrm{osc}/2$ and $T_\mathrm{osc}$. From the potential curvature near the minimum we get $T_\mathrm{osc} = 5.5$~ps. In the realistic case of our 3D nanodevice the optimal value turns out to be slightly smaller, equal $T=4.5$~ps. 

Generation of voltage pulses of a few tens of picosecond durations (picosecond pulsers\cite{pulsers1,pulsers2}) may be problematic within the current quantum technology level. 
However, the duration of the pulse applied to gates $U_{L,R}$  can be extended considerably. This results in longer conversion times but decreases error end improves the conversion fidelity. We can increase the period of oscillations $n$-times $T_\mathrm{osc}\rightarrow nT_\mathrm{osc}$ which entails a necessary decrease in voltages $V_{1..5}$  applied to the bottom gates: $V\rightarrow V/n^2$ (at the same time decreasing the necessary amplitude of $V_{0L}$ and $V_{0R}$).

\begin{figure}[b]
\includegraphics[width=7.9cm]{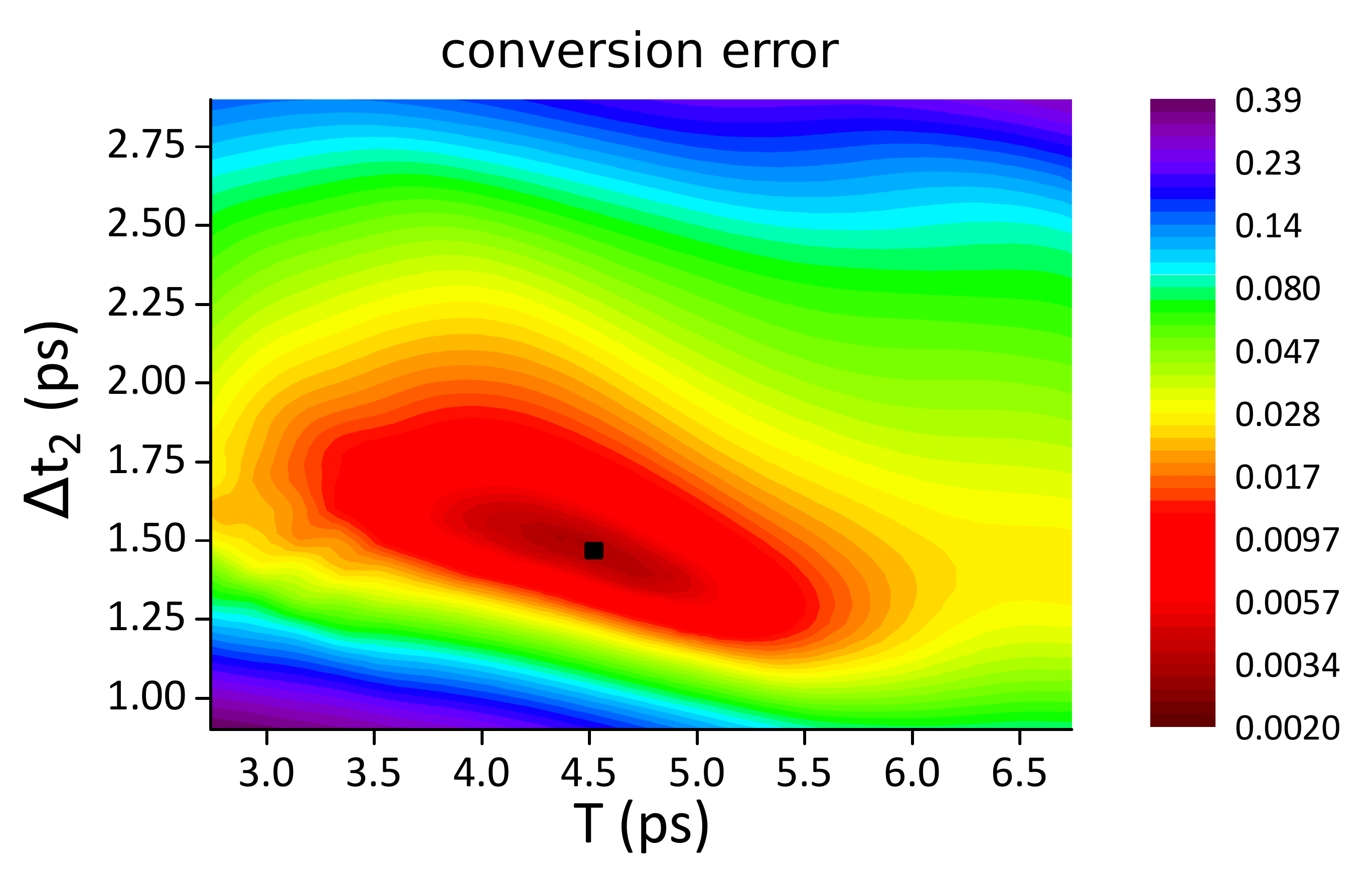}
\caption{\label{fig:7} Error of conversion for different durations $T$ of the pulse and different times of separating barrier rise $t_2=t_1+T/2+\Delta t_2$. A black square denotes optimal parameters used in a previously described simulation shown in Figs.~\ref{fig:4}-\ref{fig:6}.}
\end{figure}

A huge advantage of our solution is an appreciable error margin for parameter selection which can be seen in Fig.~\ref{fig:7}. Assuming a moment of the barrier start at $t_2=t_1+T/2+\Delta t_2$ (the pulse start at $t_1$) we plot an error map for different configurations of parameters $(T,\Delta t_2)$. We see that the error margin is indeed relatively high and to obtain fidelity ($= 1 -\mathrm{error}$) near $99$\% we have tolerance of $\pm20$\% for $T$ and $\pm5$\% for $t_2$.
And like in the analytic calculation, we obtain relatively high immunity against mismatched tuning of $T$.
To further enhance fidelity and tolerance, in comparison to Fig.~\ref{fig:7}, we have to increase the duration of the pulses and lower the voltages. To obtain fidelity of the order of $99.9\%$ the pulse duration must be at least equal $15\,\mathrm{ps}$.
The obtained picosecond time of conversion is several orders of magnitude faster than standard single-spin readout techniques based on spin-selective tunneling to a lead\cite{nowyloss}. The conversion time is also much shorter than estimated spin qubit coherence times $\sim 10$~ns for InSb nanowires\cite{10ns}.	

\section{Conclusions}
A spin-to-charge conversion mechanism is presented.
In a nanowire, in a gate-defined quantum dot, we trap a single electron of unknown spin. By applying a pulse of voltages to lateral gates we induce a Rashba spin-orbit interaction pulse. This pulse induces spatial separation of charge corresponding to spin-up and spin-down components. We then set up a potential barrier stabilizing the wavepackets in both halves of the wire.
The amount of charge in the right (and left) side of the wire is equal to the probability of finding the electron with spin-up (down) in its initial state. We thus obtain a nanodevice which effectively works as an ultrafast spin-to-charge converter. If our converter is supplemented with a charge state detector, such as a quantum point contact, we finally obtain a device performing spin state readout. In the presented configuration we obtained fidelities of the order of $99.7$\% - $99$\% for a wide range of parameters. The fidelity can be increased by applying a longer and weaker Rashba spin-orbit coupling pulse, which however increases the conversion time.


\appendix
 
\section{Materials parameters}
In Fig.~\ref{fig:1}  we show the structure of the nanodevice with the materials used. We decided to use an InSb nanowire as this material features one of the strongest Rashba spin-orbit couplings among semiconductors and the technology of catalytic growth of these nanowires is already mature. Gate-induced quantum dots have been successfully created in such structures \cite{np0,nodrr}. Thanks to strong spin-orbit coupling only one pulse of the Rashba coupling (induced by the electric field) is needed to attain sufficient spin-density separation, which significantly simplifies operation of the nanodevice. If a material with weaker spin-orbit coupling was used, multiple Rashba coupling pulses would be necessary to attain the same result, effectively increasing the readout time. The insulator material was taken as suggested in experimental works\cite{phd}. Si$_3$N$_4$ separating gates $U_{1..5}$ and $U_{L,R}$ from the nanowire was chosen to ensure a sufficiently high potential barrier at the nanowire-insulator interface (we assume a barrier of $\sim1$~eV, clearly visible in Fig.~\ref{fig:2}(middle and right)) and minimize the leakage current from the wire. A layer of SiO$_2$, however, separates gates $U_{1..5}$ from the strongly doped Si substrate which serves as a device global backgate.
For the materials used we assume the following (relative) permittivity values: $\varepsilon_\mathrm{InSb} = 16.5$, $\varepsilon_\mathrm{Si_3N_4}=7.5$, and $\varepsilon_\mathrm{SiO_2} = 3.9$,
obtaining this way space-dependent permittivity $\varepsilon(\mathbf{r})$.

\section{Numerical method: time-dependent self-consistent Schr\"odinger-Poisson calculations}
The time-dependent Schr\"{o}dinger equation is solved iteratively on a grid encompassing the nanodevice. The grid dimensions are visible in Fig.~\ref{fig:1}. The next moment of time is obtained from previous ones according to the iterative scheme:
\begin{equation}\label{eq:a1}
\Psi(\mathbf{r},t+dt)=\Psi(\mathbf{r},t-dt)+\frac{2idt}{\hbar}H(\mathbf{r},t)\Psi(\mathbf{r},t).
\end{equation}
The Hamiltonian $H(\mathbf{r},t)$, defined in Eq.~(\ref{ham}), includes the time-dependent potential $\phi(\mathbf{r},t)$. While, the spin-orbit term $H_R(\mathbf{r},t)$ of (\ref{ham}) is induced by the electric field $\mathbf{E}(\mathbf{r},t)=-\boldsymbol\nabla\phi(\mathbf{r},t)$.
The potential is calculated from the generalized Poisson equation:
\begin{equation}\label{eq:a2}
\boldsymbol\nabla\cdot\left(\varepsilon_0 \varepsilon(\mathbf{r})\boldsymbol\nabla \phi_{tot}(\mathbf{r},t)\right)=|e|\rho_e(\mathbf{r},t). 
\end{equation}
To avoid electron self-interaction we must subtract the potential generated by the electron itself
\begin{equation} 
\phi_{e}(\mathbf{r},t)=\frac{-|e|}{4\pi\varepsilon_\mathrm{InSb}\varepsilon_0}\int d^3r'\frac{\rho_e(\mathbf{r'},t)}{|\mathbf{r}-\mathbf{r'}|}
\end{equation}
from the total potential $\phi_{tot}$, thus obtaining: $\phi(\mathbf{r},t)=\phi_{tot}(\mathbf{r},t)-\phi_{e}(\mathbf{r},t)$.
Note that Eq. (\ref{eq:a2}) contains the current (at time $t$) electron density $\rho_e(\mathbf{r},t)=|\Psi(\mathbf{r},t)|^2$ calculated from the solution of the Schr\"{o}dinger equation (\ref{eq:a1}). We apply proper boundary conditions for the Poisson equation (\ref{eq:a2}) at the border of the computational grid and at the gates. To the gates we apply time-dependent potentials $V_{L,R}(t)$ and $V_{0..5}(t)$. Eq.~(\ref{eq:a2}) also takes into account the spatially varying permittivity $\varepsilon(\mathbf{r})$. Thanks to this setup we account for charge induced on the surfaces of conductors (i.e. gates) and interfaces between dielectrics. This way we obtain a realistically modeled confinement potential inside the wire presented in Fig.~\ref{fig:2}, which is then inserted into the Eq.~(\ref{eq:a1}). We thus see that the Schr\"{o}dinger equation depends on the potential obtained from Poisson equation, which in turn depends on the electron density calculated from the Schr\"{o}dinger equation and the time-dependent control voltages on the gates. Thanks to this, both equations (\ref{eq:a1} \& \ref{eq:a2}) are solved self-consistently at every time step of the nanodevice evolution. The presented method was successfully used in our previous models of quantum nanowires \cite{pawlowski1,koty1}.

\begin{acknowledgments}
This work has been supported by National Science Centre (NSC), Poland, under Grant No. 2016/20/S/ST3/00141. SB, GS and MG acknowledge support from NSC, Grant No. 2014/13/B/ST3/04526 and by Ministry of Science and Higher Education within the AGH University of Science and Technology statutory activity task no. 11.11.220.01.
This research was supported in part by PL-Grid Infrastructure.


\end{acknowledgments}


\bibliography{bibliography} 

\end{document}